\documentclass[twocolumn,prl]{revtex4}
\usepackage{graphicx}
\usepackage{dcolumn}
\usepackage{bm}
\usepackage{natbib}

\renewcommand{\(}{\left(}
\renewcommand{\)}{\right)}

\newcommand{\eq}[1]{Eq.\ (\ref{#1})}
\newcommand{\fig}[1]{Fig.\ \ref{#1}}
\newcommand{\tab}[1]{Table \ref{#1}}

\renewcommand{\deg}{\ensuremath{^\circ}}

\begin{document}


\title{Electron spin coherence in Si/SiGe quantum wells}

\author{J.\ L.\ Truitt}
 \email{jltruitt@wisc.edu}
\author{K.\ A.\ Slinker}
\author{K.\ L.\ M.\ Lewis}
\author{D.\ E.\ Savage}
\author{Charles Tahan}
\author{L.\ J.\ Klein}
\author{Robert Joynt}
\author{M.\ G.\ Lagally}
\author{D.\ W.\ van der Weide}
\author{S.\ N.\ Coppersmith}
\author{M.\ A.\ Eriksson}\email{maeriksson@wisc.edu}
\affiliation{University of Wisconsin - Madison, Madison, WI 53706}
\author{A.\ M.\ Tyryshkin}
\affiliation{Princeton University, Princeton, New Jersey 08544}
\author{J.\ O.\ Chu}
\author{P.\ M.\ Mooney}
\affiliation{IBM Research Division, T.\ J.\ Watson Research Center,
  Yorktown Heights, NY 10598}

\date{\today}

\begin{abstract}
  The mechanisms limiting the spin coherence time of electrons are of
  great importance for spintronics.  We present electron spin
  resonance (ESR) and transport measurements of six different two
  dimensional electron gases in silicon/silicon-germanium (Si/SiGe
  2DEGs).  The spin decoherence time $T_2^*$ is presented in
  conjunction with the 2DEG density $n_e$ and momentum scattering time
  $\tau_p$ as measured from transport experiments.  A pronounced
  dependence of $T_2^*$ on the orientation of the applied magnetic
  field with respect to 2DEG layer is found which is not consistent
  with that expected from any mechanism described in the literature.
\end{abstract}

\maketitle

Spin decoherence mechanisms are of fundamental importance to
spintronics.  Silicon is an excellent model system for studies of
decoherence, and electron spins in silicon have long coherence times,
making them particularly attractive for applications.  The electron
spin coherence time $T_2$ for phosphorus-bound donor electrons in
isotopically pure $^{28}$Si has been measured to be as long as 14 ms
at 7K, and extrapolates to on the order of 60 ms for an isolated
spin \cite{lyon-29si}. Loss and DiVincenzo \cite{loss} proposed the use
of spins of single electrons as quantum bits, and Kane \cite{kane} has
discussed the advantages of working in silicon.  Vrijen and
Yablonovitch et al.\ \cite{vrijen} have extended that approach to
include donor bound electrons in silicon-germanium heterostructures,
and schemes have been proposed for electron spin-based quantum
computation in silicon-germanium electron quantum dots \cite{levy,uw1}.

There have been only a few studies examining the spin coherence time of
silicon/silicon-germanium two dimensional electron gases (Si/SiGe
2DEGs) \cite{lyon-2DEG,wilamowski-prb1,wilamowski-prb2,wil-apl}. In this
paper we provide a comprehensive treatment of six different samples, using
transport measurements to extract the electron density and scattering time
and ESR to measure $T_2^*$ and to provide an indication of the spin
decoherence mechanism.  A detailed study indicates that the dominant
decoherence mechanism is strongly dependent on the orientation of the
magnetic field, but it is inconsistent with the published mechanisms.

The Si/SiGe heterostructures are grown by ultrahigh vacuum chemical
vapor deposition at the University of Wisconsin - Madison and at
IBM-Watson \cite{ibm1}.  The 2DEG sits near the top of a strained Si
layer grown on a strain-relaxed Si$_{1-x}$Ge$_x$ buffer layer, as
shown in Figure 1(a) of reference \cite{levi}.  Above the 2DEG is a
Si$_{1-x}$Ge$_x$ offset layer, followed by a phosphorus-doped dopant
layer, and then a Si$_{1-x}$Ge$_x$ spacer layer capped with Si at the
surface.  \tab{table} contains details for each sample.

\begin{table*}
\begin{tabular}{|l||c|c|c|c|c|c||c|c|c||c|c|c|}\hline
Sample & 
   \begin{tabular}{c}Si well\\(nm)\end{tabular} & 
   $x$ & 
   \begin{tabular}{c}offset\\(nm)\end{tabular} & 
   \begin{tabular}{c}dopants\\(nm)\end{tabular} & 
   \begin{tabular}{c}spacer\\(nm)\end{tabular} & 
   \begin{tabular}{c}cap\\(nm)\end{tabular} &
   \begin{tabular}{c}$n_e$\\$(10^{11}$ cm$^{-2}$)\end{tabular}& 
   \begin{tabular}{c}$\mu$\\(cm$^2$/Vs)\end{tabular} &
   \begin{tabular}{c}$\tau_{\rm p}$\\ (ps)\end{tabular} &
   \begin{tabular}{c}$T_2^*$\\$(\mu{\rm s})$\end{tabular} &
   \begin{tabular}{c}$A(15\deg)$\\{\rm anisotropy}\end{tabular} &
   $b$ \\\hline
ibm-01    & 8.0 & 0.30 & 14 & 1 & 14 & 3.5 & 4.0 & 37,300 & 4.3 & 0.6
& 1.0 & 1.6 \\
uw-030827 & 10 & 0.35 & 15 & 22 & 35 & 10 & 4.8 & 90,000 & 9.7 & 0.1 
& 4.7 & 38 \\ 
uw-030903 & 10 & 0.25 & 13 & 17 & 35 & 10 & 4.3 & 86,700 & 9.4 & 0.2 
& 2.1 & 13 \\ 
uw-031121 & 10 & 0.30 & 20 & 6 & 60 & 20 & 5.4 & 38,000 & 5.0 & 0.1
& 2.0 & 25 \\ 
uw-031124 & 10 & 0.30 & 20 & 26 & 40 & 20 & 4.7 & 63,200 & 6.9 & 0.1 
& 2.0 & 18 \\  
uw-031203 & 10 & 0.30 & 60 & 6 & 60 & 20 & 2.6 & 17,100 & 1.8 & 0.5 
& 2.3 & 10 \\\hline 
\end{tabular}
\caption{Sample parameters and measurements.  The first section of the
       table contains growth parameters.  The next section contains
       results from transport (Hall) measurements: in this section
       $\tau_p$ is calculated from the mobility $\mu$ by the
       expression $\tau_p=m_e^*\mu/e$.  The last three columns contain
       ESR measurements: $T_2^*$ is derived from the peak
       width using \eq{Hpp} ($g=2.00$ for all samples), $A(15\deg)$ is
       the anisotropy at $\theta=15\deg$ for each sample, and $b$ is
       the measured quadratic coefficient (from \eq{b}).} 
\label{table}
\end{table*}


Hall measurements are performed on all samples.  Hall bars are etch
defined and Ohmic contacts are made to the 2DEG by Au/Sb metal
evaporation and annealing at 400\deg C for 10 minutes.  
These data are used to extract the electron density and mobility, and
from the mobility we derive the momentum relaxation time $\tau_p$, an
important parameter in spin relaxation via spin-orbit and related
interactions.  The parameters reported in \tab{table} have been
corrected for a small parallel conduction path using the method of
Kane et al.\ \footnote{The unchanging slope of the transverse
  resistance shows that the conductivity of the parallel conduction
  path is much less than the conductivity of the 2DEG, and allows us
  to take a limit of Kane's expressions and extract the 2DEG mobility
  and electron density as well as the parallel conduction path
  conductivity.}, and in each case this correction was smaller than
1\% \cite{mjkane}.

Electron spin resonance data were acquired with a Bruker ESP300E X-band
spectrometer, using an Oxford Instruments ESR900 continuous flow
cryostat to maintain a sample temperature of 
4.2K.  Magnetic field calibration and tracking was done with an ER035M
NMR Gaussmeter.  

\begin{figure}
\centerline{\includegraphics[width=0.3\textwidth]{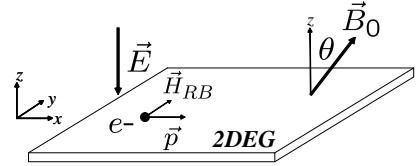}}
\caption{Electrons in the quantum well move in the presence of an
  electric field between the dopants and the well, and therefore feel
  an effective in-plane magnetic field.}
\label{rashba}
\end{figure}

\begin{figure*}
\centerline{\includegraphics[width=0.75\textwidth]{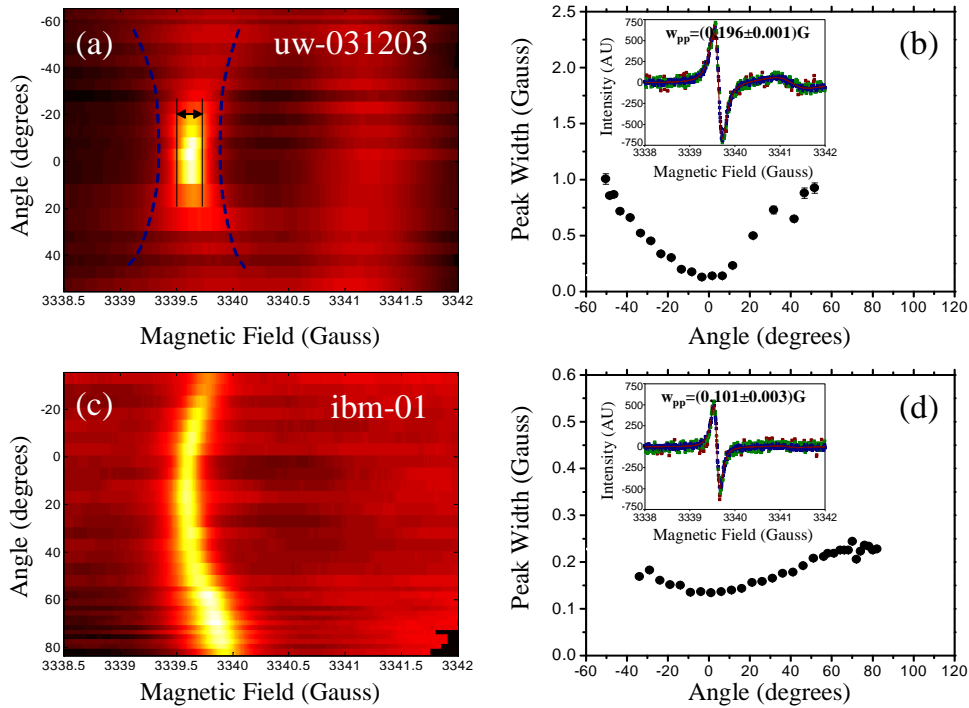}}
\caption{
  Orientation map of the ESR signal from (a) sample uw-031203 and (c)
  sample ibm-01 where the $x$-axis is the magnetic field, the shading
  scale is the peak intensity, and the $y$-axis is the orientation
  angle.  Lorentzian-fit peak width (see inset) of (b) sample
  uw-031203 and (d) sample ibm-01 as a function of orientation angle.}
\label{031203+ibm6min}
\end{figure*}

The ESR spectra for all samples were measured as a function of the
orientation of the applied magnetic field, determined by the angle
$\theta$ between the magnetic field and the growth direction of the
sample (see \fig{rashba}).  Figures \ref{031203+ibm6min}(a) and (c)
are two-dimensional maps of the ESR intensity as a function of
magnetic field and orientation angle for two selected samples.  The
peak-to-peak ESR linewidths $\Delta H_{pp}$ were extracted by fitting
the lineshapes to the derivative of a Lorentzian (see insets in
\fig{031203+ibm6min}(b) and (d)).  These linewidths show a pronounced
dependence on the orientation angle $\theta$ (see
\fig{031203+ibm6min}(b) and (d)).  The minimal ESR linewidth (at
$\theta=0$) and the observed linewidth anisotropy for all samples are
summarized in Table \ref{table}.\footnote{In many of the ESR data sets
  (e.g., the inset of \fig{031203+ibm6min}(b)) there is a small
  peak near 3341 Gauss (in the region of Land\'e g-factor $g\approx
  2.0$) with no orientational dependence that is wider than the 2DEG
  peak.  Because this peak is almost perfectly equidistant between two
  42G split phosphorous peaks (not shown in the figures), we ascribe
  this peak to electrons in the dopant layer shared among clusters of
  phosphorous nuclei (see reference \cite{feher}, Figures 15 and 16).}

The ESR linewidth $\Delta H_{pp}$ is directly related to the coherence time
$T_2^*$ through the expression \cite{poole}
\begin{equation}
 \Delta H_{pp}={\frac{2}{\sqrt{3}}}{\frac{\hbar}{g\mu_B}}
   \({\frac{1}{T_2^*}}\),
\label{Hpp}
\end{equation}

\noindent where $g$ is the Land\'e g-factor and $\mu_B$ is the Bohr
magneton.  It has been proposed \cite{wilamowski-prb2} that the
orientation dependence of $T_2^*$ (and thus of $\Delta H_{pp}$) in
similar 2DEG structures results from a D'yakonov-Perel spin relaxation
mechanism due to fluctuating Rashba fields \cite{DP}. Electrons in the
quantum well leave behind positive charge on their donors, setting up
an electric field that is in addition to any interface electric field.
The electrons in the well on the Fermi surface move with the Fermi
velocity in this electric field and therefore feel an effective
magnetic field in the plane of the 2DEG (see \fig{rashba}).  This
field is called the Rashba field $H_{RB}$.  Scattering of the 2D
electrons results in a fluctuating field $H_{RB}$ which is always in
the 2DEG plane (see \fig{rashba}).  Therefore, when the external
magnetic field $B_0$ is applied perpendicular to the 2DEG
($\theta=0$), the fluctuating $H_{BR}$ are perpendicular to $B_0$.
However, when $B_0$ is tilted with respect to the 2DEG ($\theta\ne
0$), a component of the fluctuating field appears along $B_0$ driving
decoherence more rapidly, which results in an orientational dependence
of $T_2^*$ and the ESR linewidth $\Delta H_{pp}$ through \eq{Hpp}.  In
general, there may be other contributions to the linewidth as well,
and the spin coherence time $T_2^*$ can be written as

\[ {\frac{1}{T_2^*}}={\frac{1}{T_2^{BR}}}+{\frac{1}{T_2'}}, \]

\noindent where $1/T_2^{BR}$ is the Rashba term contribution, and $1/T_2'$
includes all other contributions to the linewidth (such as
inhomogeneous broadening or other decoherence mechanisms).

Two groups have derived expressions for $T_2^{BR}$ in the limit
$\omega_c\tau_p\cos\theta\gg 1$.  Both can be written as 
\begin{eqnarray}
{\frac{1}{T_2^{BR}}}=\alpha^2k_F^2\tau_p\left[ 
    {\frac{\eta}{1+(\omega_c\cos\theta)^2\tau_p^2}}\sin^2\theta
  \right.\nonumber\\\left.
  + {\frac{1/2}{1+(\omega_L-\omega_c\cos\theta)^2\tau_p^2}}
    \(\cos^2\theta+1\)\right],
\label{wil}
\end{eqnarray}

\noindent where $\eta=1/2$ for \cite{wilamowski-prb2} and $\eta=2$ for
\cite{tahanwc}, $\alpha$ is the Rashba coefficient (defined by the
Hamiltonian $\mathcal{H}=\alpha(\sigma\times{\bf k}_F)\cdot\hat{n}$,
$\sigma$ are the Pauli spin matrices), ${\bf k}$ ($k_F$) is the
electron (Fermi) wavevector, $\tau_p$ is the momentum relaxation time,
$\theta$ is the angle the magnetic field makes relative to the sample
growth direction, $\omega_c=eB/m_e^*$ is the cyclotron frequency, and
$\omega_L=g\mu_BH/\hbar$ is the spin precession (Larmor)
frequency \cite{wilamowski-prb2}. The limit
$\omega_c\cos\theta\tau_p\gg 1$ implies that \eq{wil} is valid only
for small angles $\theta$.

If $1/T_2^{RB}$ is the dominant term in $1/T_2^*$, then Equation
\ref{wil} can be normalized as follows to obtain an expression that
does not depend on the value of the Rashba parameter
$\alpha$,\footnote{The presumed origin of the Rashba field in these
  samples is charge left behind in the dopant layer and the breakdown
  of the effective mass approximation at the sharp quantum well
  interface.  In general, any charge asymmetry can give rise to an
  electric field that would drive decoherence in the same manner.
  There are four different types of asymmetries: (a) bulk inversion
  asymmetry (BIA) due to the unit cell of the growth materials
  \cite{pfeffer}, (b) structural inversion asymmetry (SIA) due to
  growth structure (e.g., the location of dopants)
  \cite{pfeffer}, (c) native interface asymmetry (NIA) due to chemical
  bonding at the interface \cite{olesberg}, and (d) fluctuations in
  the dopant concentration \cite{shermanAPL}. Neither (a) BIA nor (c)
  NIA are present in Si/SiGe heterostructures \cite{shermanPRB}, which
  leaves (b) SIA and (d) dopant concentration fluctuation as the
  possible sources of electric fields.  In our case, with an
  asymmetrically doped Si/SiGe quantum well, the SIA will come from
  the dopant layer as in (b) and (local) fluctuations in the charge
  density (d).  Together with interface effects, these will influence
  the value of $\alpha$.} but only on the momentum scattering time
$\tau_p$:
\begin{eqnarray}
 A(\theta)&\equiv&{\frac{\Delta H_{pp}(\theta)}{\Delta H_{pp}(0)}}=
{\frac{1/T_2^*(\theta)}{1/T_2^*(0)}} \nonumber\\
&=&\left[1+\(\omega_L-\omega_c\)^2\tau_p^2\right]\nonumber\\
&& \times\left[ {\frac{\eta}{1+(\omega_c\cos\theta)^2\tau_p^2}}\sin^2\theta
  \right.\nonumber\\
&& \left.+ {\frac{1/2}{1+(\omega_L-\omega_c\cos\theta)^2\tau_p^2}}
    \(\cos^2\theta+1\)\right].
\label{anisotropy}
\end{eqnarray}

\begin{figure}
\centerline{\includegraphics[width=0.5\textwidth]{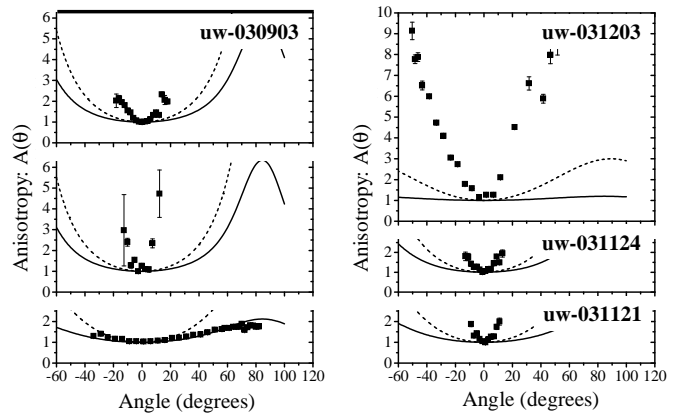}}
\caption{Anisotropy (normalized peak width) as a function of
  orientation angle for all samples, and \eq{anisotropy} (dashed
  $\eta=2$, solid $\eta=1/2$) for each using $\tau_p$ as reported in
  Table \ref{table}.}
\label{poorFits}
\end{figure}

\begin{figure}
\centerline{\includegraphics[width=0.5\textwidth]{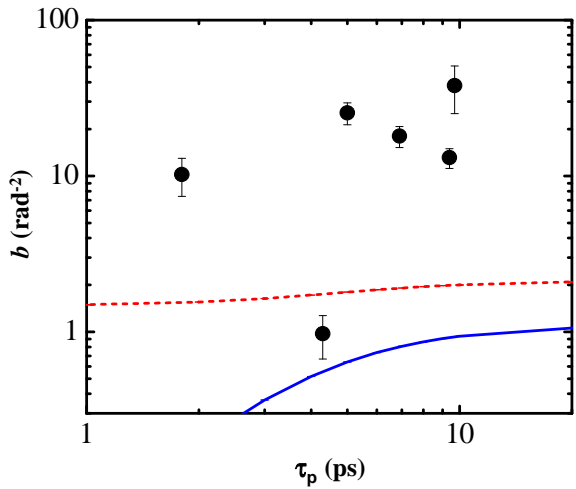}}
\caption{The quadratic coefficient $b$ of \eq{anisotropy} (dashed
  $\eta=2$, solid $\eta=1/2$) near the origin as a function of the
  parameter $\tau_p$, including the measured sample $b$'s at transport
  $\tau_p$'s (see Table \ref{table}).}
\label{b}
\end{figure}

The orientational dependence of the normalized ESR linewidths for all
samples and \eq{anisotropy} (using each sample's transport-measured
$\tau_p$) are plotted in \fig{poorFits}.  The observed anisotropies at
small angles differ substantially from those that \eq{anisotropy}
predicts.  To see this quantitatively, and since this expression only
applies for small $\theta$, we Taylor expand \eq{anisotropy} to
\begin{equation}\label{parabola} 
A(\theta)=1+b\theta^2,
\end{equation}

\noindent where the quadratic coefficient $b$ is a measure of how
quickly the anisotropy increases with angle $\theta$.  For each sample
the quadratic coefficient $b$ can be extracted from a parabolic fit
(of \eq{parabola}) to the data.  A plot of the quadratic coefficient
$b$ as a function of the parameter $\tau_p$ with the measured samples
is given in \fig{b} for both values of $\eta$.  For all six samples
the quadratic coefficients $b$ differ substantially from both
theoretical predictions.  Even more striking, the maximum quadratic
coefficient $b$ that \eq{anisotropy} can give (for any value of
$\tau_p$) is 1.07 rad$^{-2}$, which is nearly an order of magnitude
smaller than that observed for five of the six samples.
 

As \fig{b} shows for these samples, the semi-classically derived
$1/T_2^{BR}$ does not account for the observed $1/T_2^*$.  A fully
quantum mechanical derivation of the Rashba decoherence $1/T_2^{BR}$
may be necessary, or there may be other mechanisms contributing to the
linewidth through the additional component $1/T_2'$.  If the latter is
the case, then $1/T_2'$ is necessarily orientationally-dependent,
since an orientationally-independent term cannot change the functional
form of the anisotropy, and so cannot affect $b$.  That is, the
observed discrepancy must be due to an orientationally-dependent
effect.  Since silicon possesses an inverison symmetry,
orientationally dependent mechanisms originating from the
anti-symmetric term in the Hamiltonian introduced by
Dresselhaus \cite{dresselhaus} (see references \cite{ivchenko} and
\cite{wilamowski-prb2}) should not contribute to the linewidth.  It is
conceivable that the data are showing some kind of orientationally
dependent {\it inhomogeneous} broadening.  One way to test this would
be to do pulsed EPR experiments, measuring $T_2$ instead of $T_2^*$
and removing the sensitivity to inhomogeneous broadening (such as the
static dipole-dipole interactions with residual $^{29}$Si).  The
interactions to residual $^{29}$Si can also be eliminated through the
use of isotopically pure $^{28}$Si in the quantum well.


In summary, we have performed a combination of ESR and transport
measurements on six Si/SiGe 2DEGs and characterized the orientation
dependence of the ESR linewidths.  We observe an orientationally
dependent spin decoherence with an anisotropy larger than any current
theory predicts.

This work was supported in part by the NSA and ARDA under ARO contract
number W911NF-04-1-0389, and by the National Science Foundation under
Grant Nos.\ DMR-0325634 and DMR-0079983.

\bibliography{refs}

\end{document}